# Use of multiple covariates in assessing treatment-effect modifiers: A methodological review of individual participant data meta-analyses

Peter J Godolphin[1], Nadine Marlin[2], Chantelle Cornett[1], David J Fisher[1], Jayne F Tierney[1], Ian R White[1], Ewelina Rogozińska[1]

**Corresponding Author:** Peter J Godolphin, 2nd Floor, 90 High Holborn, MRC Clinical Trials Unit at University College London, Institute of Clinical Trials and Methodology, London, WC1V 6LJ, UK. p.godolphin@ucl.ac.uk

**Affiliations**

1: MRC Clinical Trials Unit at University College London, Institute of Clinical Trials and Methodology, London, UK

2: Pragmatic Clinical Trials Unit, Barts and The London School of Medicine and Dentistry, Queen Mary University of London, London, UK

**ORCID ID's**

Peter Godolphin (0000-0003-0648-0992); Nadine Marlin (0000-0002-2841-4825); David Fisher (0000-0002-2512-2296); Jayne Tierney (0000-0002-4734-3014); Ian White (0000-0002-6718-7661); Ewelina Rogozińska (0000-0003-3455-0644)

# Abstract

Individual participant data (IPD) meta-analyses of randomised trials are considered a reliable way to assess participant-level treatment effect modifiers but may not make the best use of the available data. Traditionally, effect modifiers are explored one covariate at a time, which gives rise to the possibility that evidence of treatment-covariate interaction may be due to confounding from a different, related covariate. We aimed to evaluate current practice when estimating treatment-covariate interactions in IPD meta-analysis, specifically focusing on involvement of additional covariates in the models. We reviewed 100 IPD meta-analyses of randomised trials, published between 2015 and 2020, that assessed at least one treatment-covariate interaction. We identified four approaches to handling additional covariates: (1) Single interaction model (unadjusted): No additional covariates included (57/100 studies); (2) Single interaction model (adjusted): Adjustment for the main effect of at least one additional covariate (35/100); (3) Multiple interactions model: Adjustment for at least one two-way interaction between treatment and an additional covariate (3/100); and (4) Three-way interaction model: Three-way interaction formed between treatment, the additional covariate and the potential effect modifier (5/100). IPD is not being utilised to its fullest extent. In an exemplar dataset, we demonstrate how these approaches can lead to different conclusions. Researchers should adjust for additional covariates when estimating interactions in IPD meta-analysis providing they adjust their main effects, which is already widely recommended. Further, they should consider whether more complex approaches could provide better information on who might benefit most from treatments, improving patient choice and treatment policy and practice.

**Keywords:** treatment-covariate interaction, effect modification, participant-level covariate, confounding, meta-analysis, individual participant data

# 1. Introduction

Individual participant data (IPD) meta-analysis of randomised trials is recognised as the most reliable and flexible way to assess participant-level treatment effect modifiers and enables researchers to make the best use of the available data[1]. Identifying whether and how treatment effects vary across different participant groups (referred to as an interaction) is vital to informing how best to treat individual patients[2-4] and thus to improving patient outcomes.

Traditionally, meta-analysis methods have pooled results across trials for each patient subgroup and compared the subgroup meta-analyses results using a chi-square test of interaction[5]. As this approach is at risk of aggregation bias[5-8], an alternative approach has recently been proposed, in which interactions are first estimated within each trial and then pooled using a standard meta-analysis model[6, 9]. However, both approaches usually consider only one-covariate at a time. Thus, estimates of interaction derived from a one-covariate at a time approach may be due, in part at least, to the interaction effect of a different covariate. This could be seen as a form of confounding. For example, if a cancer treatment is effective only in metastatic disease, and older patients are more likely to have metastatic disease, then an interaction between treatment and age would be expected. This would be a genuine interaction, because treatment would on average benefit older patients more than younger patients; but it would be a confounded interaction, because the interaction with metastatic disease provides a fuller and more clinically useful description.

Meta-analyses dealing with several related participant-level covariates may be at the biggest risk of such confounding occurring. In IPD meta-analysis, access to participant-level data enables additional covariates to be incorporated when estimating treatment-covariate interactions, which may alleviate the impact of potential participant-level confounding. These covariates could be included in several ways, from simple adjustment for their main effects, which is highly recommended[4], through to forming higher-level interaction terms that may in fact alter the question that the interaction analysis aims to answer. Currently, it is unclear if, and how researchers utilise this opportunity and what the impact of their chosen approach is.

Thus, the aim of this article is to describe and critique current practice when estimating treatment-covariate interactions in an IPD meta-analysis, specifically focusing on involvement of additional covariates in the models. The article is structured as follows. In Section 2, we outline the methodology of the literature review. Section 3 describes the findings of the review, detailing the snapshot of current practice in IPD meta-analysis when estimating treatment-covariate interactions. In Section 4, we present example statistical models for each identified approach, the question each model addresses and apply these approaches to an exemplar dataset for illustration purposes. We follow with a discussion and brief conclusions.

## 2. Review methods

### 2.1 Protocol

Our literature review followed a prospectively registered protocol[10] (version from 2nd August 2021) that shares some methods, including the search strategy, with a review by Marlin *et al.*[11,12] (PROSPERO no. CRD42019126768).

### 2.2 Literature search

Using a cohort of IPD meta-analysis reports published between 2015 and 2020[11], we included IPD meta-analyses with at least two randomised trials in which at least one participant-level treatment-covariate interaction was reported. Articles were screened for eligibility by one reviewer (NM), with another reviewer (PJG, ER, CC) independently confirming eligibility.

### 2.3 Sample size

We sought to identify 100 reports of eligible IPD meta-analyses. A random sample from the potentially eligible records was obtained by selecting each 10$^{th}$ record until we either reached the predefined sample size or exhausted the number of potentially eligible records.

### 2.4 Data collection and extraction

We used a bespoke data collection form that was piloted on five eligible studies (see Appendix S1). The form was split into four sections: general information (publication year, medical area, number of trials included, IPD meta-analysis approach), non-linear effects (only relevant for the related review[11]), effect modification (number of treatment-covariate interactions assessed, number of outcomes considered for effect modification, whether additional covariates were involved beyond the treatment-covariate interaction in any statistical models, and if so how they were selected/involved) and contact details of the IPD meta-analysis team.

Data were extracted in duplicate by at least two reviewers (two from PJG, ER, CC, NM), with discrepancies resolved through discussion. Extracted data was collated in an Excel spreadsheet and cross-checked for accuracy against the completed data extraction forms.

We sought additional documents for eligible IPD meta-analyses (i.e., protocol, statistical analysis plan etc.) to supplement the data extraction, as appropriate. Where further clarifications were required, we contacted one of the authors of the IPD meta-analysis for more details. For pragmatic reasons, we also made several assumptions when extracting data. These included:

1. A one-stage fixed-effect approach had been used if the methods described an analysis "stratified by trial" or "adjusted for trial"[13, 14].
2. The software used was indicative of the analytical approach. For example, if Stata commands metan or ipdmetan[15] had been used we assumed a two-stage approach.
3. A random-effects model had been used if the methods mentioned "a random effect for study/trial" or "random effects model". Conversely, a fixed-effect model was assumed if the methods described a "random intercept".
4. We assumed that no covariates (over and above the effect modifier itself) were involved if none were mentioned in either the methods or in tables/figures displaying the results of the statistical models.

**2.5 Data synthesis and analysis**

All data in this review are summarised narratively. Continuous variables are presented with mean and standard deviation or median and interquartile range. Categorical variables are described with frequency counts and percentages. All analyses were performed using Stata software (version 16.1).

# 3. Results of methodological review

In order to reach 100 IPD meta-analyses meeting out eligibility criteria, we had to assess a random sample of 211 from the 738 potentially eligible records (Figure 1).

**Figure 1:** Study selection flow diagram

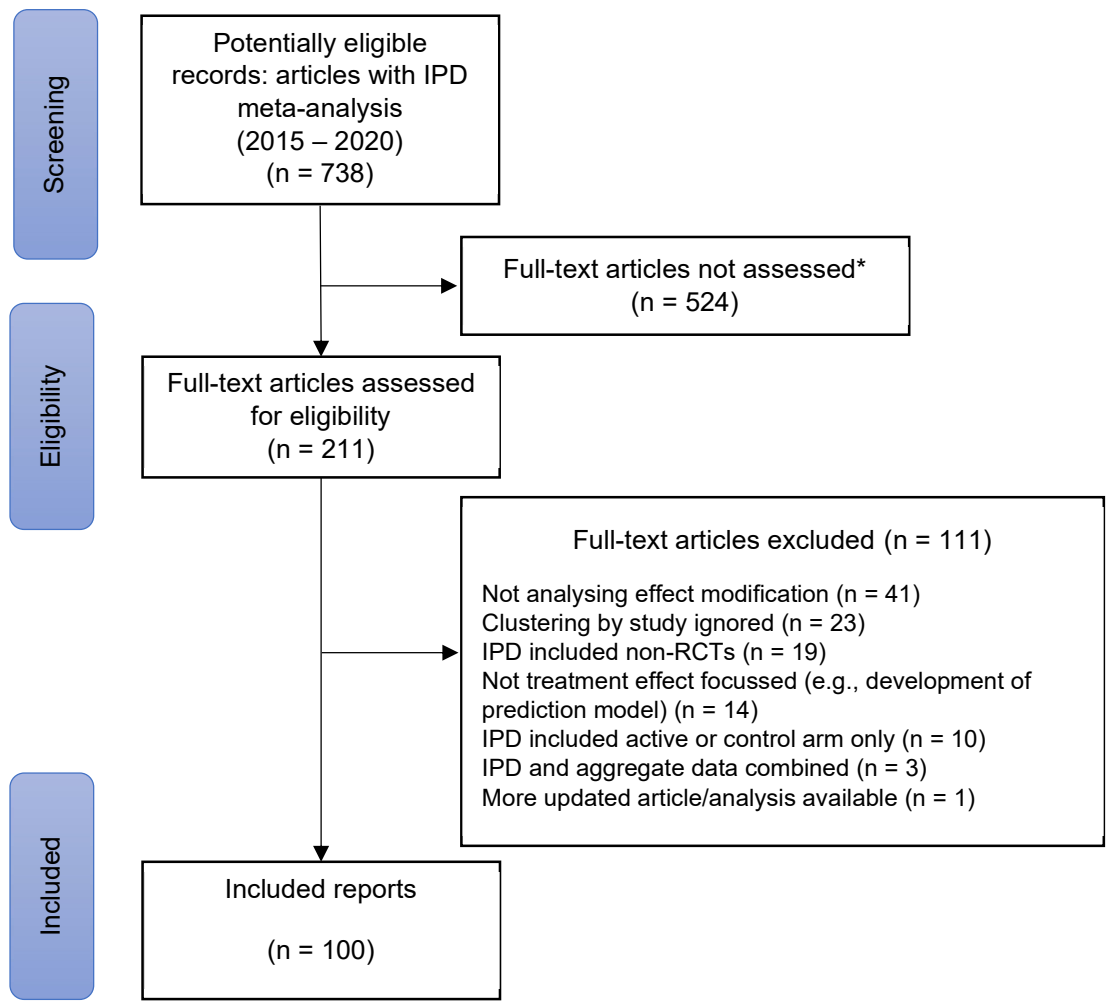

*Full-text articles were not assessed for eligibility once 100 eligible IPD meta-analyses had been identified
IPD refers to individual participant data, RCT refers to randomised controlled trial

The included reports of IPD meta-analyses are evenly distributed in terms of their publication year between 2015 and 2020 (Table 1). They comprised a variety of medical fields, with 29% addressing questions in cardiovascular research. About a quarter of the studies were prospectively registered on PROSPERO. Median meta-analysis size was five datasets (trials) and 2816 participants. Most of meta-analyses used a one-stage IPD method either as the sole approach (60%) or in addition to a two-stage approach (19%). A two-stage IPD method as the only approach was rare (19%). Fixed-effect models were used slightly more than a random-effects model (39% vs 30%), although a fifth of studies used both fixed-effect and random-effects models, with one often employed as a sensitivity analysis.

**Table 1:** Characteristics of the included individual participant data meta-analyses

|  |  | Total (n=100) |
|---|---|---|
| **Year published** | | |
| | 2015 | 14 (14%) |
| | 2016 | 8 (8%) |
| | 2017 | 19 (19%) |
| | 2018 | 19 (19%) |
| | 2019 | 21 (21%) |
| | 2020 | 19 (19%) |
| **Medical field** | | |
| | Cardiovascular | 29 (29%) |
| | Neurology | 13 (13%) |
| | Cancer | 12 (12%) |
| | Women's Health | 7 (7%) |
| | Mental Health | 7 (7%) |
| | Critical Care | 5 (5%) |
| | Public Health | 4 (4%) |
| | Infectious Diseases | 3 (3%) |
| | Neonatal Health | 3 (3%) |
| | Other[1] | 17 (17%) |
| **PROSPERO registration** | | |
| | Yes | 22 (22%) |
| | Not reported | 78 (78%) |
| **Number of trials[2]** | | |
| | Median [25th, 75th] | 5 [3, 11] |
| | Min, Max | 2, 34 |
| **Number of participants included[3]** | | |
| | Median [25th, 75th] | 2816 [1094, 4754] |
| | Min, Max | 73, 174 000 |
| **IPD synthesis method** | | |
| | One-stage | 60 (60%) |
| | Two-stage | 19 (19%) |
| | Both | 19 (19%) |
| | Unclear | 2 (2%) |
| **Statistical model(s) used** | | |
| | Fixed-effect only | 39 (39%) |
| | Random-effects only | 30 (30%) |
| | Both | 20 (20%) |
| | Unclear | 11 (11%) |

All data are frequency (%) unless stated
IPD refers to individual participant data
[1]Other category includes a variety of medical fields that are represented by two or less IPD meta-analyses (see Appendix S2)
[2]Excludes five IPD meta-analyses that included a varying number of datasets
[3]Excludes nine IPD meta-analyses where multiple numbers of participants were reported (n=7) or no information was reported (n=2)

Effect modification tended to be explored based on a small number of outcomes (median of two), although this ranged from 1 to 16 (Table 2). A median of six covariates per IPD meta-analysis were explored as potential effect modifiers. A quarter of studies (26%) treated continuous covariates as continuous when estimating treatment-covariate interactions while in almost two-thirds (63%) continuous covariates were categorised.

Only around a third of studies (35%) provided sufficient detail to ascertain whether within- and across-trial information was appropriately separated out when estimating treatment-covariate interactions, either written in the main text, appendices or in supplementary documents (e.g., in protocols or statistical analysis plans) (Table 2). Of these, 15 used methods that separated out within- and across-trial information. Less than half (43/100) of the included IPD meta-analyses involved additional covariates when estimating treatment-covariate interactions.

**Table 2:** Description of investigation into effect modification in the included sample of individual participant data meta-analyses

|  |  | Total (n=100) |
|---|---|---|
| **Number of outcomes investigated for effect modification** | | |
| | Median [25th, 75th] | 2 [1,3] |
| | Min, Max | 1, 16 |
| **Number of effect modifiers considered** | | |
| | Median [25th, 75th] | 6 [2, 9] |
| | Min, Max | 1, 28 |
| **Covariate type of effect modifiers in each study[1]** | | |
| | Categorical | 81 (81%) |
| | Continuous | 26 (26%) |
| | Categorised continuous | 63 (63%) |
| **Handling of within- and across-trial information in analysis of effect modification** | | |
| | Separated out[2] | 15 (15%) |
| | One-stage, unclear if separated out | 43 (43%) |
| | Conflated[2] | 20 (20%) |
| | Unclear | 23 (23%) |
| **Involvement of additional covariates when assessing effect modification** | | |
| | No | 57 (57%) |
| | Yes | 43 (43%) |

All data are frequency (%) unless stated
[1]Data are not mutually exclusive
[2]One study carried out one analysis that correctly separated out within- and across-trial information and another that conflated within- and across-trial information

In IPD meta-analyses that involved additional covariates, most of these covariates were selected *a priori* with only three studies reporting use of a stepwise procedure (Table 3). The reasons for involving additional covariates varied across the studies (Table 3). They were mostly included as main effects alongside the treatment-covariate interaction of interest (35/43, 81%). In eight studies, the covariates were used in a more complex model: adjustment for at least one additional two-way treatment-covariate interaction (3/43, 7%) or inclusion of a three-way interaction between two covariates and treatment (5/43, 12%).

**Table 3:** Description of involvement of additional covariates when assessing effect modification in the 43 individual participant data meta-analyses where this occurred

| Item | Total (n=43) |
|---|---|
| **How were additional covariates selected** | |
| Determined *a priori* | 28 (65%) |
| Stepwise procedure | 3 (7%) |
| Not mentioned/unclear | 12 (28%) |
| **How were covariate(s) involved when assessing effect modification** | |
| Main effect of covariate(s) adjusted for | 35 (81%) |
| Two-way interaction with covariate adjusted for | 3 (7%) |
| Three-way interaction formed with treatment, effect modifier and covariate[1] | 5 (12%) |
| **Reasons given for involving additional covariate(s)** | |
| *Confounding explicitly mentioned* | 5 (12%) |
| *Confounding not explicitly mentioned* | 28 (65%) |
| Known prognostic factors | 13 (30%) |
| To account for baseline covariates/baseline imbalance | 10 (23%) |
| Based on previous research | 3 (7%) |
| Stepwise selection | 2 (5%) |
| *No reasons given* | 10 (23%) |

All data are frequency (%)
[1]One study stratified by a covariate and then assessed effect modification within these strata, which is indirectly forming a three-way interaction with the covariate, the effect modifier and treatment and is included in this category

# 4. Approaches to estimating treatment-covariate interactions in the context of inclusion of additional covariates

In our review we identified four different approaches to handling additional covariates when estimating treatment-covariate interactions, with most of the studies estimating interactions without including additional covariates. The identified approaches are as follows:

**Approach 1**: *Single interaction model (unadjusted).* When estimating the treatment-covariate interaction the analysis does not involve any additional covariates and is unadjusted.

**Approach 2**: *Single interaction model (adjusted).* When estimating the treatment-covariate interaction the analysis is adjusted for the main effect of at least one additional covariate.

**Approach 3:** *Multiple interactions model.* When estimating the treatment-covariate interaction the analysis adjusts for at least one two-way interaction between an additional covariate (often a further potential effect modifier) and treatment.

**Approach 4:** *Three-way interaction model.* When estimating the treatment-covariate interaction the analysis includes a three-way interaction term with treatment, the potential effect modifier, and an additional covariate (often a further potential effect modifier).

Table 4 contains example statistical models for the four approaches for both one-stage and two-stage meta-analysis, with these models presented for a continuous outcome and correctly separating out within- and across-trial information[4, 8]. The approaches 2-4 (Table 4) assume only involvement of one additional covariate, although approaches 2 and 3 can easily be extended by making $w_{ij}$, $\beta_{2i}$ and $\gamma_2$ vectors. Approach 4 can potentially be extended to include higher-order interaction terms as appropriate. Note, we do not make a distinction here between fixed-effect and random-effect models, as this is not relevant when specifying the covariates to be included in a model.

The four identified approaches address three different questions (Table 4). Approaches 1 and 2 use different methods to address the same question in the single interaction model: this validly estimates an interaction effect, and this information could be appropriately used to inform treatment decisions. However, if there is at least one true additional effect modifier, then incorporating this in the multiple interactions model (approach 3) could lead to more nuanced treatment recommendations. With multiple potential effect modifiers, the three-way interaction model (approach 4) could be used to test whether the multiple interactions model is appropriate, or whether there is an additional relationship between the effect modifiers that needs to be accounted for. In this situation, the question addressed could be thought of as "Does the multiple interactions model describe the data adequately?". Note that for the three-way interaction model, $\gamma_1$-$\gamma_3$ could all be coefficients of interest, and in this case the question addressed is "How does the average treatment effect vary between individuals with higher and lower levels of $w$ and $z$?".

**Table 4:** Summary of the identified approaches to involving additional covariates when estimating treatment-covariate interactions in the sample of 100 individual participant data meta-analyses

| Approach | Description | Example one-stage linear mixed model | Example two-stage linear model | Question addressed |
|---|---|---|---|---|
| 1 | Single interaction model (unadjusted) | $Y_{ij} = \alpha_i + \beta_{1i}z_{ij} + \theta x_{ij} + \boldsymbol{\gamma} x_{ij}(z_{ij} - \bar{z}_i)$ | **Stage 1:** $Y_{ij} = \alpha_i + \beta_{1i}z_{ij} + \theta_i x_{ij} + \gamma_i x_{ij} z_{ij}$<br>**Stage 2:** Pool $\gamma_i$ in a meta-analysis to estimate $\boldsymbol{\gamma}$ | How does the average treatment effect vary between individuals with higher and lower levels of $z$? |
| 2 | Single interaction model (adjusted) | $Y_{ij} = \alpha_i + \beta_{1i}z_{ij} + \beta_{2i}w_{ij} + \theta x_{ij} + \boldsymbol{\gamma} x_{ij}(z_{ij} - \bar{z}_i)$ | **Stage 1:** $Y_{ij} = \alpha_i + \beta_{1i}z_{ij} + \beta_{2i}w_{ij} + \theta_i x_{ij} + \gamma_i x_{ij} z_{ij}$<br>**Stage 2:** Pool $\gamma_i$ in a meta-analysis to estimate $\boldsymbol{\gamma}$ | Same as 1, but with possibly greater power due to covariate adjustment |
| 3 | Multiple interactions model | $Y_{ij} = \alpha_i + \beta_{1i}z_{ij} + \beta_{2i}w_{ij} + \theta x_{ij} + \boldsymbol{\gamma_1} x_{ij}(z_{ij} - \bar{z}_i) + \gamma_2 x_{ij}(w_{ij} - \bar{w}_i)$ | **Stage 1:** $Y_{ij} = \alpha_i + \beta_{1i}z_{ij} + \beta_{2i}w_{ij} + \theta_i x_{ij} + \gamma_{i1} x_{ij} z_{ij} + \gamma_{i2} x_{ij} w_{ij}$<br>**Stage 2:** Pool $\gamma_{i1}$ in a meta-analysis to estimate $\boldsymbol{\gamma_1}$ | Among individuals with the same level of $w$, how does the average treatment effect vary between individuals with higher and lower levels of $z$? |
| 4 | Three-way interaction model | $Y_{ij} = \alpha_i + \beta_{1i}z_{ij} + \beta_{2i}w_{ij} + \theta x_{ij} + \gamma_1 x_{ij}(z_{ij} - \bar{z}_i) + \gamma_2 x_{ij}(w_{ij} - \bar{w}_i) + \boldsymbol{\gamma_3} x_{ij}(z_{ij} - \bar{z}_i)(w_{ij} - \bar{w}_i)$ | **Stage 1:** $Y_{ij} = \alpha_i + \beta_{1i}z_{ij} + \beta_{2i}w_{ij} + \theta_i x_{ij} + \gamma_{i1} x_{ij} z_{ij} + \gamma_{i2} x_{ij} w_{ij} + \gamma_{i3} x_{ij} z_{ij} w_{ij}$<br>**Stage 2:** Pool $\gamma_{i3}$ in a meta-analysis to estimate $\boldsymbol{\gamma_3}$ | Does the way the average treatment effect varies between individuals with higher and lower levels of $w$ itself vary with the level of $z$? |

$Y_{ij}$ is the outcome, which is continuous in these example models. $\alpha_i$ is intercept for each trial. $z_{ij}$ is the potential effect modifier, $\bar{z}_i$ bar is study specific mean of this covariate. $\theta$ is the treatment effect, with $x_{ij}$ an indicator variable for treatment (Yes, No). $w_{ij}$ is potential participant-level confounding covariate (or other potential effect modifier). $\gamma$ is the coefficient for the within-trial interaction.
The coefficient of interest is bolded for each approach (e.g., $\boldsymbol{\gamma}$).
Note that the model numbers correspond to the four approaches described in Section 4

## 4.1. Application to exemplar dataset

To demonstrate how these approaches may provide different estimates of interaction effects, we used an exemplar dataset resembling in its characteristics an average IPD meta-analysis from the sample of 100 included in our review (details in the supplement). This exemplar dataset comprises five 'trials' of varying sizes (range from 200-1500), with a total of 3200 participants and a continuous outcome (score from 0-50). Analysis was undertaken in Stata software (version 16.1).

Under a fixed-effect inverse-variance weighted meta-analysis using a two-stage model, this example shows a beneficial effect of treatment and limited statistical heterogeneity (Figure 2).

**Figure 2:** Fixed-effect meta-analysis of five trials for a continuous outcome using exemplar dataset

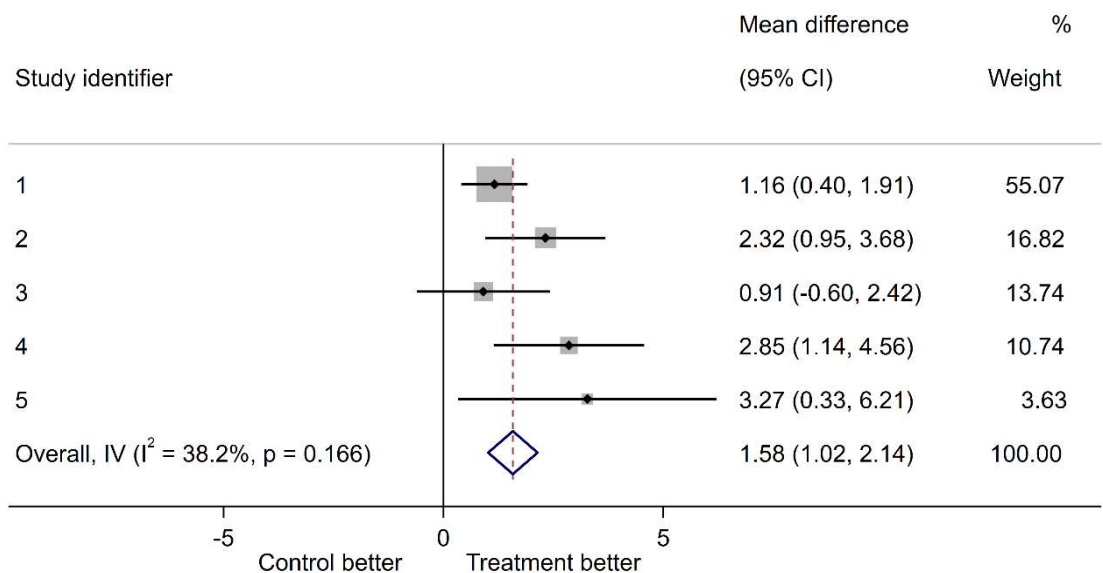

To demonstrate the methods in action, this dataset contained two binary covariates (referred to as *Covariate 1* and *Covariate 2*, with these covariates indicating some baseline characteristic such as a co-morbidity) that had a degree of negative correlation (Pearson's correlation coefficient, ρ=-0.31), with both appearing to modify the effect of treatment on their own (using the single interaction model (unadjusted), Table 4). We then fitted the single interaction model (adjusted), adjusting for the other covariate and then finally fitted the multiple interactions model. All pooled interactions (differences in mean differences due to the continuous nature of the outcome) were estimated using a fixed-effect model following the within-trials approach that correctly separates out across- and within-trial information[9]. The pooled differences in mean differences and interaction p-values for each approach and each covariate are presented in Figure 3.

**Figure 3:** Differences in mean differences and interaction p-values for the investigation of effect modifiers Covariate 1 and Covariate 2 in the exemplar dataset, dependent on model choice

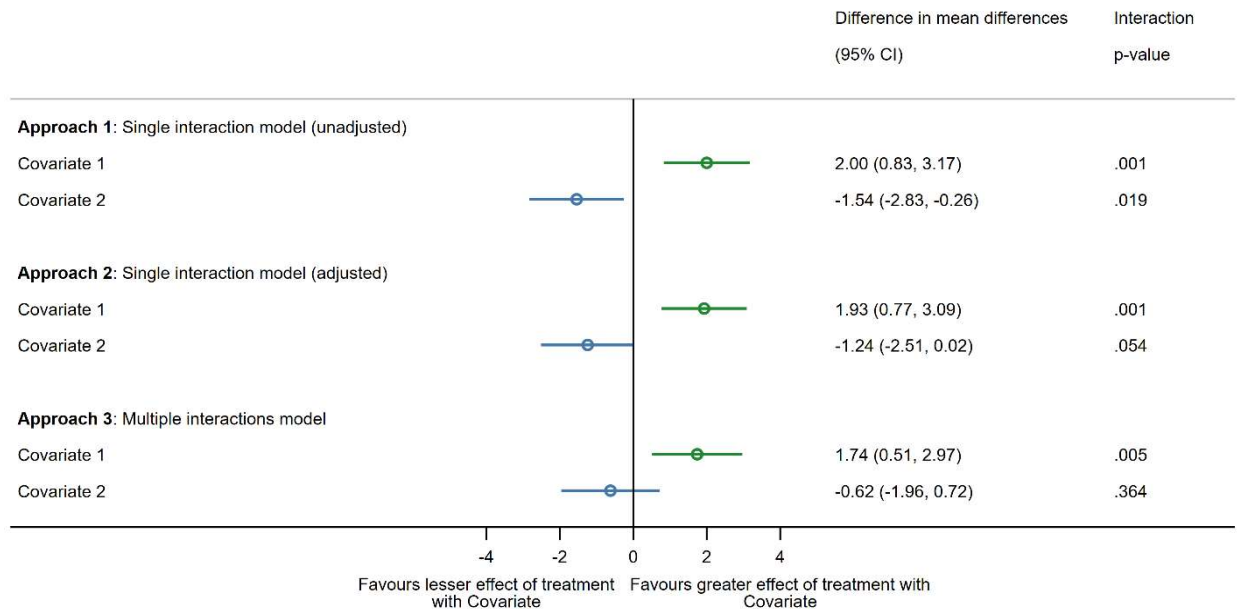

In our illustrative example, there is consistent evidence that Covariate 1 has a modifying effect on treatment, indicating that those with the Covariate (e.g., co-morbidity) have greater benefit from treatment compared to those without the Covariate (Figure 3). However, evidence of Covariate 2's modifying effect reduces as the main effect of Covariate 1 is adjusted for (Approach 2) and disappears when the two-way interaction of treatment and Covariate 1 is adjusted for (Approach 3). This indicates that the Covariate 1 by treatment interaction explains the Covariate 2 by treatment interaction. If either of the two single interaction models had been the main approach in this example, then it may have been concluded that both covariates are important when making treatment decisions, whereas the multiple interaction model demonstrates that Covariate 1 is driving this modifying effect. We found no evidence of a three-way interaction between both covariates and treatment, suggesting that the multiple interactions model describes the data adequately.

## 5. Discussion

Most IPD meta-analyses of randomised trials included in our sample did not involve additional covariates when assessing patient-level, treatment-covariate interactions. When they did, it was often to adjust main effects, rather than to adjust for additional treatment-covariate interactions or to investigate higher-order interactions between treatment and multiple covariates. Very few IPD meta-analyses included multiple treatment-covariate interactions in the same model, yet such models have potential to be most adept at accounting for participant-level confounding and produce the most appropriate estimates of treatment-covariate interactions.

Interactions are not confounded in the same way that exposure effects are confounded in observational studies. For example, using our exemplar dataset, we found evidence that the Covariate 2 by treatment interaction is confounded by the Covariate 1 by treatment interaction. This means the best way to predict treatment effects is by using Covariate 1 (as shown by the multiple interactions model), but it remains true that treating based on Covariate 2 is not invalid if Covariate 1 is not available, even if this is driven, in part, by the Covariate 1 by treatment interaction.

Whilst the availability of IPD from trials creates the possibility to fit more complex models, these can present practical problems. Incorporating multiple treatment-covariate interactions may be challenging due to model convergence issues[16]. Inclusion of more covariates increases the probability of missing covariate data. Power can be reduced through fitting multiple treatment-covariate interactions in the same model and/or modelling more complex higher-order interactions. Potential solutions to these challenges may exist, such as mean imputation of missing data at the trial level[17, 18]. However, proper evaluation of the implications that these methodological challenges pose for the results of treatment-covariate interaction analyses is warranted before widespread advocation of such methods. Of note, in our review, none of the studies used a risk-based approach as advocated by the PATH statement[19] or shrinkage approaches that incorporate multiple treatment-covariate interactions at once[20]. Both methods have been popularised only in recent years, and our sample does not reflect this newer methodology.

Within our sample, reporting on whether within- and across-trial information was separated was limited. Appropriate methodology to separate out across- and within-trial information for both one-stage and two-stage approaches exists and has been available for several years[6, 7]. Reassuringly, a greater number of studies in our review used an appropriate analysis that did not introduce aggregation bias (15%) compared to a prior review by Fisher *et al*.[6] ( 2%). However, the number of studies undertaking appropriate analyses and/or reporting clearly is still severely lacking and methodological and reporting improvements are warranted. Research should be guided by the Preferred Reporting Items for a Systematic Review and Meta-analysis of IPD[21]; which preceded many of included studies included in this review.

Our study was guided by a prospectively developed protocol however is not without limitations. First, we included a pre-determined sample of 100 IPD meta-analyses, and we did not assess an additional 200-300 IPD meta-analyses of randomised trials between 2015 and 2020. Also, some studies only investigated one treatment-covariate interaction and others may not have found evidence of multiple effect modifiers. Both factors may have impacted whether studies would have considered involving additional covariates at all, potentially reducing the number that utilised more complex approaches. However, we reviewed statistical analysis plans and protocols (where available) and rarely found consideration of involving additional covariates contained in this additional documentation. Further, whilst including multiple treatment-covariate interactions or forming three-way interactions may not have been appropriate for these cases, simple covariate adjustment (Single interaction model (adjusted)) would have been possible. Finally, we often had to make assumptions when extracting data, but we did contact authors for clarification on key information, such as reasons for involving additional covariates, where required.

Future methodological studies should establish a general strategy for estimating treatment-covariate interactions whilst taking account of other covariates. Empirical data should be analysed to understand in which circumstances the various possible approaches will lead to different conclusions and based on these findings, advocate that the most appropriate approaches should be used in future IPD meta-analyses of randomised trials. Barriers to using newer methodology, such as risk-based modelling, could be explored to identify why this has not had large take-up from the evidence synthesis community.

## 6. Conclusions

Researchers should make the most of the IPD they collect, and at a minimum adjust for prognostic factors when estimating treatment-covariate interactions, provided they adjust their main effects, which is already widely recommended. Where multiple effect modifiers are explored, consideration of additional methods that alleviate any potential participant-level confounding, such as the multiple interactions model, may provide better information on which participants are most likely to benefit from treatments, leading to more informed treatment policy and practice.

## Funding

PJG is part supported by the National Institute for Health Research (NIHR)'s Development and Skills Enhancement Award (NIHR301653). PJG and DJF are part supported by Prostate Cancer UK (https://prostatecanceruk.org/) Grant number: RIA 16-ST2-020. PJG and DJF are part supported and JFT, IRW and ER are fully supported by the UK Medical Research Council (https://mrc.ukri.org/) Grant numbers: MC_UU_00004/06. NM is funded by an NIHR Doctoral Fellowship (DRF-2018-11-ST2-077). The funders had no role in study design, data collection and analysis, decision to publish, or preparation of the manuscript.

## Data availability statement

All data that was extracted as part of the literature review and used in the exemplar dataset is available on the Open Science Framework (see https://osf.io/zmu75/).

## Conflict of interest

The authors declare that they have no competing interests

## Authors contributions

Peter J Godolphin: Conceptualization, Formal analysis, Investigation, Writing – Original Draft, Supervision

Nadine Marlin: Investigation, Resources, Writing – Review & Editing, Supervision

Chantelle Cornett: Investigation, Writing – Review & Editing

David J Fisher: Writing – Review & Editing

Jayne F Tierney: Writing – Review & Editing

Ian R White: Writing – Review & Editing

Ewelina Rogozińska: Conceptualization, Investigation, Writing – Review & Editing, Supervision

# Acknowledgments

We thank Dr Claire Vale for useful comments and suggestions for this work. Note that this acknowledgement does not necessarily imply endorsement.

# Appendicies

# Appendix S1: Data extraction form for IPD meta-analysis studies

**Section A: General information**

| Administrative | |
|---|---|
| **Date extracted** | |
| **Name of person extracting** | |
| **First author** | |
| **Research group / collaboration name** | |
| **IPD MA characteristics** | |
| **Publication Year** | |
| **Medical Field** | |
| **Number of datasets analysed** | |
| **Number of participants in those datasets** | Min:　　　　　Max: |
| **Number of participants analysed from those datasets** | Min:　　　　　Max:　　　　　Total: |
| **Dataset identification** | ☐Systematic review COCHRANE <br> ☐Literature search (*includes if an existing IPD dataset was used but updated with a literature search) <br> ☐ Data sharing platform <br> ☐ Collaboration – Prospective IPD with shared protocol <br> ☐ Collaboration – existing IPD dataset (*includes studies using data from a previously published IPD analysis, e.g. done in the same research unit) <br> ☐ Research program <br> ☐ Trial registry database <br> ☐ Patient registry database <br> ☐ Company database <br> ☐ Not reported |
| **PROSPERO ID (if given)** | |
| **Types of study data included** | ☐RCTs only <br> ☐Any other study types |
| **IPDMA approach** | ☐1 stage, fixed effects |
|  | ☐1 stage, random effects |
|  | ☐2 stage, fixed effects |
|  | ☐2 stage, random effects |
|  | ☐Trial effect ignored |
| **General modelling approach** |  |
| **Data format of primary outcome** | ☐Binary |
|  | ☐Categorical |
|  | ☐Continuous |
|  | ☐Time to event |
| **Is any outcome reported a composite?** | ☐Yes |
|  | ☐No |
| **Sample size considerations** | ☐Power calculation |
|  | ☐Post hoc power assessment |
|  | ☐Limited to general talk of "IPDs" have larger power compared to aggregate Metanalysis or single trials |
|  | ☐None (*includes if discussion mentions that the study may not have had enough power to detect an effect size) |
| **Is multiple testing accounted for?** | ☐Yes |
|  | ☐No |
|  | ☐Not mentioned |
| **Complex associations** | ☐Non-linear effects – GO TO SECTION B |
|  | ☐Effect modification – GO TO SECTION C |
|  | ☐Subgroup analysis without comparison test |
|  | ☐Neither |
|  | *NOTE: effect modification is not considered between treatment effect and trial. This is usually done to test homogeneity and does not involve interest in identification of effect modifiers.* |

## Section B: Complete only for papers assessing non-linear effects

|  |  |
|---|---|
| **Analysis approach for investigation of non-linear trends** |  |
| **Justification for analysis method** |  |

**Section C: Complete only for papers assessing effect modification**

| Type of covariate in the effect modification (tick multiple) | ☐Continuous variable (e.g. age, weight) <br> ☐Categorised continuous variable (e.g. age <35 vs age 35+) <br> ☐Categorical variable (e.g. gender, marital status) |
|---|---|
| Number of effect modifications considered in total | |
| Number of outcomes looked at for effect modification | |
| Analysis approach for investigation of effect modification | |
| Are additional covariates included beyond the treatment-covariate interaction for any effect modification? | |
| If so, how are these covariates selected? | ☐Univariate analysis <br> ☐Stepwise procedure <br> ☐Determined a priori <br> ☐Other, please describe…………………………………………………… |
| If so, how are these covariates included? | ☐Additional variables included in the model <br> ☐3-way interaction (2 covariates interacted with treatment) <br> ☐Multiple 2-way interactions included in the model <br> ☐A higher-order factorial covariate constructed from two other categorical covariates <br> ☐Other, please describe…………………………………………………… |
| If so, are reasons reported? What are they? | |

**Section D: Contact details of IPD meta-analysis team**

| Corresponding authors name, email, and job/study role | |
|---|---|

| Principal investigator's name and email. If corresponding author is principal investigator then leave blank | |
| Lead statistician's name and email. If corresponding author is lead statistician, then leave blank | |

## Appendix S2: Other medical fields (from Table 1)

**Table S1:** Medical fields that occurred in two of less IPD meta-analyses

|  | Total (n=17) |
|---|---|
| Addiction/Alcoholism | 1 (6%) |
| Complementary and alternative medicine | 2 (12%) |
| Dentistry | 1 (6%) |
| Gastroenterology | 1 (6%) |
| Infection | 1 (6%) |
| Nephrology | 2 (12%) |
| Nutrition | 2 (12%) |
| Respiratory | 2 (12%) |
| Rheumatology | 1 (6%) |
| Sleep Medicine | 1 (6%) |
| Surgery | 2 (12%) |
| Virology | 1 (6%) |

All data are frequency (%)
IPD refers to individual participant data